# Simulation-Based Study of AI-Assisted Channel Adaptation in UAV-Enabled Cellular Networks

Andrii Grekhov[1*], hrekhov.andrii@npp.kai.edu.ua, ORCID: 0000-0001-7685-8706,

Volodymyr Kharchenko[2], volodymyr.kharchenko@npp.kai.edu.ua, ORCID: 0000-0001-7575-4366

Vasyl Kondratiuk[3], vasyl.kondratiuk@kai.edu.ua, ORCID: 0000-0002-5690-8873

[1, 2*, 3] Research Training Center "Aerospace Center",
State University "Kyiv Aviation Institute", Kyiv, 1, Liubomyra Huzara ave, Kyiv, 03058, Ukraine.

*Corresponding author: [1*] hrekhov.andrii@npp.kai.edu.ua , +380 98 286 57 30

**Abstract**

This paper presents a simulation-based study of Artificial Intelligence (AI)-assisted communication channel adaptation in Unmanned Aerial Vehicle (UAV)-enabled cellular networks. The considered system model includes communication channel "Ground Base Station – Aerial Repeater – UAV Base Station – Cluster of Cellular Network Users". The primary objective of the study is to investigate the impact of adaptive channel parameter control on communication performance under dynamically changing interference conditions.

A lightweight supervised machine learning approach based on linear regression is employed to implement cognitive channel adaptation. The AI model operates on packet-level performance indicators and enables real-time adjustment of Transaction Size (*TS*) in response to variations in Bit Error Rate (*BER*) and effective *Data Rate*. A custom simulation environment is developed to generate training and testing datasets and to evaluate system behavior under both static and adaptive channel configurations.

The performance of the proposed AI-assisted adaptation scheme is assessed using *Latency*, channel Average Utilization (*AU*), and packet transmission characteristics as key metrics. Simulation results demonstrate that adaptive control of transaction size improves latency stability and channel utilization compared to a static configuration, particularly in high-interference scenarios. The findings confirm that even computationally efficient AI models can provide measurable performance benefits for UAV-assisted cellular communication systems and are suitable for implementation on resource-constrained UAV platforms.

**Keywords:** UAV-assisted cellular networks, AI-assisted channel adaptation, simulation modeling, latency, average channel utilization, BER

## 1. Introduction

In recent years, UAV-enabled communication systems integrated with terrestrial cellular networks (Figure 1) have attracted significant research interest due to their ability to provide flexible and rapidly deployable connectivity [1–4]. UAVs can operate as aerial relays or airborne base stations, extending cellular coverage in remote areas or in scenarios where ground infrastructure is unavailable, damaged, or overloaded [5].

The integration of UAVs into cellular networks enables new communication paradigms characterized by high mobility, dynamic channel conditions, and varying interference levels [6–10]. Artificial intelligence and machine learning techniques are increasingly applied in such systems to support adaptive resource management, channel estimation, and performance prediction. However, many existing studies focus on routing, positioning, or spectrum allocation, while relatively less attention is paid to cognitive adaptation of packet-level channel parameters under dynamically changing interference conditions.

In parallel, clustering of ground users has been shown to improve scalability and reduce channel contention in UAV-assisted cellular networks [11–15]. By serving groups of users rather than individual terminals, UAV-based base stations can reduce signaling overhead and improve channel utilization, particularly in dense or highly dynamic environments.

This paper adopts a simulation-based approach to study AI-assisted adaptation of UAV communication channel parameters in a clustered cellular network topology. The considered communication chain includes a ground base station, an aerial repeater UAV, a UAV base station, and a cluster of cellular network users. The focus of the study is not on proposing a new network architecture, but on analyzing how adaptive control of transaction size based on channel conditions affects latency and average channel utilization.

The main contributions of this work are as follows:
- development of a simulation model for a UAV-enabled cellular communication chain with clustered users;
- analysis of packet latency and average channel utilization as functions of transaction size, data rate, and bit error rate;
- implementation of a lightweight AI-assisted channel adaptation algorithm based on supervised learning.

The remainder of the paper is organized as follows. Section 2 reviews related work. Section 3 presents the system model and simulation scenario, along with analytical and simulation results. Section 4 describes the AI-assisted adaptive data transfer mechanism. Conclusions and future research directions are provided in Section 5.

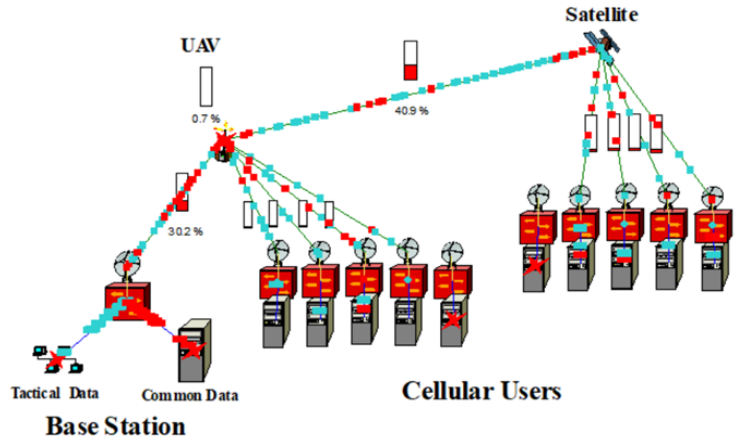

**Figure 1**: Example of UAV-enabled cellular network

## 2. Related Works

The article [1] considers the main paradigms in UAV cellular communications: UAV-to-UAV cellular communications as new aerial users served by ground base stations, and UAV-to-UAV cellular communications as new aerial communication platforms serving ground users. Results are obtained for the optimal UAV flight trajectory, based on which methods are proposed for finding approximate solutions to the trajectory using graph theory and convex optimization methods.

A multi-UAV wireless communication system in which multiple aerial base stations mounted on UAVs are used to serve a group of users on the ground is considered in [2]. To obtain fair access among users, the minimum throughput for all ground users was maximized by optimizing the scheduling and association of multi-user communication in conjunction with the trajectory and power control of UAVs.

A UAV-enabled wireless network in which multiple UAVs are deployed as airborne base stations to serve ground users is considered in [3]. The total throughput for all ground users in the downlink is maximized by optimizing the UAV deployment locations together with the bandwidth and power allocation of both access and return links. Alternating optimization and sequential convex programming techniques are used to obtain a locally optimal solution. Simulations show that the proposed scheme significantly improves the total throughput among all ground users compared to other benchmark schemes.

The user-level and cellular network performance that serves both UAVs and ground users in the downlink is studied in [4]. The advantageous signal propagation conditions that UAVs receive due to their altitude often work against them. This is due to the increased co-channel interference received from neighboring ground base stations, which is not compensated for by the improved signal strength. Compared to a ground user in an urban area, the analysis shows that a UAV flying at an altitude of 100 m can experience a 10x reduction in throughput and a drop in coverage from 76% to 30%. It is shown that depending on the UAV altitude and its antenna configuration, the performance of an aerial user can scale better with network density than that of a ground user.

The review [5] presents the latest UAV communication technologies, task modules, antennas, resource processing platforms and network architectures. Methods such as machine learning and path planning for improving existing UAV communication methods are considered. Encryption and optimization methods for communication and for power management are discussed. Applications of UAV networks for navigation and surveillance, ultra-reliable and low-latency communications, edge computing and AI are given. The main issue of the review is the interaction between UAVs, advanced cellular communications and IoT.

AI is currently developing rapidly and is very successful due to the large amount of data available. Therefore, the integration of intelligence into the core of UAV networks has begun, using AI algorithms to solve a number of problems associated with drones. The article [6] provides an overview of some applications of AI in UAV-based networks.

The objective of the paper [7] is to review AI-based autonomous UAV networks. The classification of autonomous functions, network resource management and scheduling, multiple access and routing protocols, as well as power management and energy efficiency for UAV networks are considered. It is shown that AI-based UAVs are a technologically feasible and economically viable paradigm for next-generation autonomous networks.

Intelligent resource allocation using artificial ecosystem optimizer with deep learning technology (IRA-AEODL) in UAV networks is presented in [8]. This technology aims to efficiently allocate resources in wireless UAV networks. The IRA-AEODL method focuses on maximizing the system utility for users, user association, energy planning and trajectory design.

The problem of joint UAV-user association, channel allocation, and transmission power control in a UAV network with multi-connectivity support with limited feedback is considered in [9]. The objective is to mitigate co-channel interference while maximizing the long-term utility of the system. The problem is modeled as a cooperative stochastic game with a hybrid discrete-continuous action space. To solve this problem, the Multi-Agent Hybrid Deep Reinforcement Learning (MAHDRL) algorithm is proposed.

In the paper [10], UAV with wireless communication and edge computing support is studied. The UAV equipped with radio frequency chains and servers can sustainably provide wireless energy to charge IoT devices and perform computing tasks from these devices while hovering at designated hover points. The goal is to minimize the weighted sum of energy consumption and information age in this system, which depends on the hover time of the UAV at designated points and its flight time. To achieve this, the deployment of hover points and the order in which the UAV visits these points are jointly optimized.

A systematic study of practical Ultra Reliable and Low Latency Communications (URLLC) procedures in IoT was carried out in review [11]. The selected methods are divided into four categories: structure-based, diversity-based, metaheuristic-based, and channel state-based. The advantages and disadvantages of using URLLC in IoT network are outlined.

The review [12] provides an overview of the integration of UAVs into cellular networks. In particular, types of consumer UAVs, interference issues, standardization solutions, serving air users with existing ground base stations, capabilities of flying repeaters and UAV-based base stations, prototype test benches, new rules for commercial use of UAVs, and cellular security in using UAVs.

The review [13] examines air-to-ground, ground-to-ground, and air-to-air channels for UAV communications in various cases. Recommendations are provided for managing the UAV communication link budget, taking into account losses and channel attenuation effects. The gains from transmit/receive diversity and spatial multiplexing when using multiple antennas are analyzed. Issues for modeling UAV communication channels are discussed.

The article [14] provides guidance on the use of UAVs in wireless communications. 3D UAV deployment, performance analysis, channel modeling and energy efficiency are studied. Open problems and research directions in UAV communications are presented. Optimization theory, machine learning, stochastic geometry, transportation theory, and game theory are covered. The manual contains recommendations for the analysis, optimization and design of wireless communication systems based on UAVs.

In the paper [15], data collection for mass Machine-to-Machine Communication (mMTC) networks supported by UAV stations moving in the air is studied. Machine-to-Machine Communication Device (MTCD) data collection, which can be achieved by different approaches, is important for the operation of mMTC networks. In a generalized model, where the target MTCDs are grouped into several clusters, a UAV station moves across the clusters and collects data from each cluster by hovering over the cluster. The corresponding MTCD clustering strategy, UAV hovering strategy, and UAV flight strategy affect the energy consumption of the system. MTCD clustering is performed using the Greedy Learning Clustering (GLC) algorithm. A modeling method based on the idea of Artificial Energy Map (AEM) is proposed to find the optimal hovering position in the cluster.

An energy-efficient user clustering mechanism based on Gaussian mixture models using a modified expected maximization algorithm is proposed in [16]. The algorithm is designed to provide initial user clustering and drone deployment. The proposed algorithm improves the system energy efficiency by 25% and the connection reliability by 18.3% compared with other baseline methods.

A wireless network with Macro Base Station (MBS) and cellular users is considered in [17]. A new wireless network topology is proposed that supports UAVs, which are deployed at the cell edge to serve edge users with poor communication quality. To avoid large interference, the locations of these UAVs are modeled as a

Homogeneous Poisson Point Process (HPPP) under a Poisson Cluster Distance Constraint (PCDC). Edge users are clustered around each UAV, and their locations are modeled as Poisson Cluster Processes (PCP). The Laplace transforms of intra-cluster interference, inter-cluster interference, and other interference are derived. Subsequently, the coverage probability and area spectrum efficiency are derived for UAVs and MBS using stochastic geometry tools.

The energy efficiency of a new system model is analyzed [18] where UAVs are used to cover user access points (user clusters) and are deployed at charging pads to increase the flight time. A new concept of "cluster pairs" is introduced to capture the dynamic nature of user spatial distribution. Using stochastic geometry tools, a new distance distribution is derived, which is vital for the energy efficiency analysis. The coverage probability is calculated under two deployment strategies: (i) one UAV per cluster pair and (ii) one UAV per cluster. The energy efficiency for both strategies is calculated. Numerical results show which of the two strategies is better for various system parameters. New aspects of UAV-based communication system such as dynamic user density and deployment strategies of one or two UAVs per cluster pair are investigated. It is shown that there is an optimal cluster pair density to maximize the energy efficiency.

Wireless communication technologies face limitations in capacity and coverage range. In the paper [19], UAV clusters are proposed as a solution to provide wireless network services to ground users to solve this problem. At the same time, the quality of service of cellular networks is optimized and the wireless coverage range is increased. The model uses an algorithm for optimizing the coverage of the UAV cluster for the swarm optimization scenario. Simulation experiments are conducted on the proposed algorithm, and the simulation results show that the algorithm has good convergence.

UAVs have become carriers of aerial Base Stations (BS) to cover temporary hotspots and serve mobile users. Using UAV-BS for rapid deployment and offloading faces the problem of time-efficient joint optimization of multiple objectives, such as the number of UAV-BS swarms, three-dimensional deployment locations, and user distribution. In the paper [20], the joint optimization problem is transformed into a combinatorial problem and solved based on data that has both time efficiency and robustness. A generative framework based on neural network is proposed. The results show that the proposed method solves the problem with fewer computation iterations compared with benchmark methods.

In the study [21], content-oriented Device-to-Device (D2D) communication of UAVs is considered. Clustering of D2D users (i.e., ground users) is considered and UAV delivers only the requested content to the cluster head nodes. The clustering approach is considered since the objective is to reduce the energy consumption of UAV during the communication phase. Clustering of ground nodes will allow UAV to communicate only with the cluster head nodes compared to a larger group of users. The cluster head nodes are then responsible for forwarding the cached content to the appropriate cluster members. The evaluation of various performance parameters such as throughput, energy consumption, and content delivery latency yielded promising results across all parameters.

The objective of the paper [22] is to develop a technique called "Fuzzy UAV Path" that supports smooth UAV path design and enables ground network topology change. A comprehensive UAV-based data collection model is proposed to achieve dynamic orchestration of wireless ground sensors. A software-defined wireless sensor network is implemented in the ground network that adaptively supports UAV motion and improves communication energy efficiency. An analytical delay analysis is proposed across network orchestration phases, which provides flexible range for UAV path design. Various simulation tools, such as MATLAB, CupCarbon, Contiki-Cooja, and Mission Planner, are used for modeling and performance evaluation.

Report [23] examines the problem of protecting UAV communication channels from adaptive AI jamming. A countermeasure model is proposed in which the defensive AI onboard UAV uses Q-Learning for dynamic frequency switching, while the attacking AI employs a transition prediction model. A simulation is conducted and it is shown that the RL agent ensures stable communication even with aggressive attacker adaptation.

3.  **System Model and Simulation Scenario**

Our model (Figure 2) of the communication channel consists of Ground Base Station, Aerial Repeater at a distance of 10 km from it, UAV at a distance of 10 km from Repeater. UAV acts as an air base station for the ground cellular network. The ground network has N users, the number of which varies from one to five and are located at a distance of 1 km from the air base station. Transactions of 10, 100, 1000, 100000 bits are transmitted to users of the ground cellular network at *Data Rates* of 6, 10, 45 Mbps.

The simulation focuses on packet-level performance metrics rather than specific physical or MAC-layer protocols, making the results applicable to a wide range of cellular technologies.

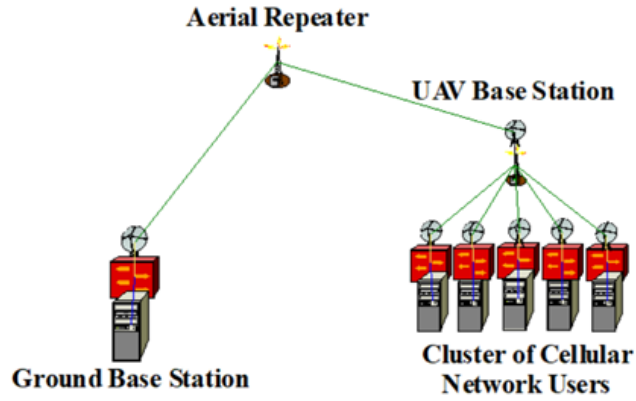

**Figure 2**: Model "Ground Base Station – Aerial Repeater – UAV Base Station – Cluster of Cellular Network Users"

### 3.1 Dependences of *Latency* on Transaction Size

The total packet latency is defined as the sum of transmission delays over each link in the communication chain:

$$L_{total} = L_{BS\ to\ Repeater} + L_{Repeater\ to\ UAV} + L_{UAV\ to\ User}, \quad (1)$$

where $L = TS/Data\ Rate$. Since the Data Rate is the same at each section, the latency at each section is calculated the same way, but taking into account the distance between the channel elements. Code for Latency calculation is given in Appendix 1.

For each link, the transmission delay is calculated as the ratio of transaction size to data rate. Assuming identical data rates for all links, the total latency is proportional to the transaction size and inversely proportional to the data rate.

Figure 3 illustrates the dependence of latency on transaction size for different data rates. As expected, latency increases with transaction size, while higher data rates result in lower overall delay. These results confirm the suitability of the model for estimating packet delays in UAV-assisted multi-hop communication scenarios.

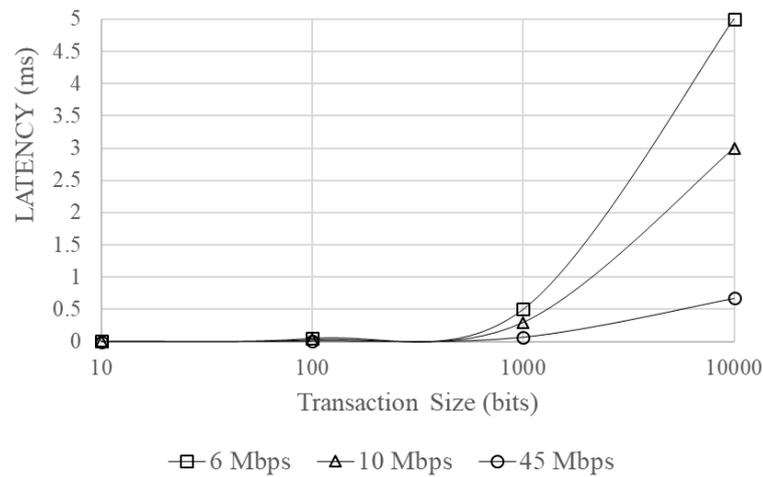

**Figure 3**: Dependences of *Latency* on *TS* for different *Data Rates* (N = 1)

### 3.2. Dependences of "Ground Base Station - Aerial Repeater" channel *Utilization* on Transaction Size

Channel *Utilization* is determined by the formula:

$$Utilization = (TS \times N/(Data\ Rate \times T)) \times 100\%, \quad (2)$$

where *T* is the transmission time. Program code for Utilization calculation is given in Appendix 2 and dependences of channel "Ground Base Station - Aerial Repeater" *Utilization* on *TS* for *Data Rate* = 1.54 Mbps are given in Figure 4.

The graphs show how the channel *Utilization* increases with the packet size and the number of users. The X-axis is on a logarithmic scale for better visualization of the range of packet values. The larger the packet size and the number of users, the higher the channel *Utilization*, which brings it closer to the maximum channel capacity. The resulting graphs and tables allow you to visually assess the impact of packet size and the number of users on channel *Utilization* and help find optimal parameters for stable data transmission without channel overload.

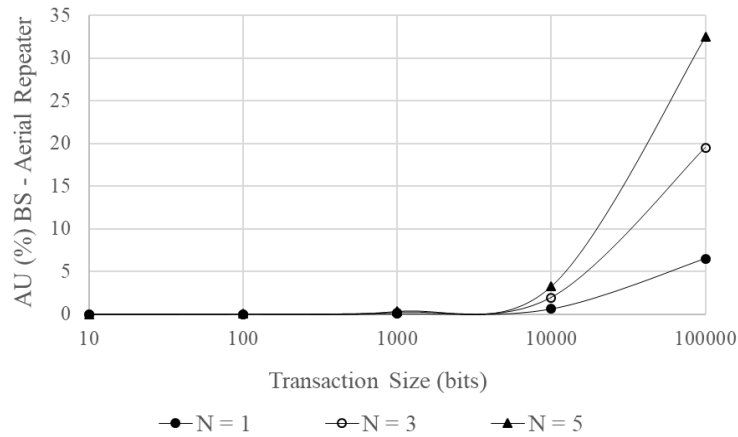

**Figure 4:** Dependences of "Ground Base Station - Aerial Repeater" channel *Utilization* on *TS* for *Data Rate* = 1.54 Mbps and *T* = 1 s

### 3.3. Dependences of "Ground Base Station - Aerial Repeater" channel *Utilization* on *Data Rate*

Let us consider the case when 100 Kbits packets are transmitted from a ground base station to users of a ground network at rates of 1.544, 2.048, 4, 6, 10, 34, 45 Mbps. To find graphs of the dependences for *Utilization* on the data transfer rate we will use the formula from the previous paragraph. The program code for *Utilization* calculation in this case is shown in Appendix 3, and the dependences of channel "Ground Base Station - Aerial Repeater" *Utilization* on *TS* for different *Data Rates* are shown in Figure 5.

The graphs show how the channel Utilization changes as the data rate increases. At low speeds, the channel load is higher because the throughput is lower. As the data rate increases, the channel load decreases because the data is transferred faster.

The results help to estimate how much the channel is loaded at different transmission rates. For example, high load at low transmission rates can lead to delays or congestion, so such data is useful for optimizing transmission parameters and choosing appropriate speeds for efficient network operation.

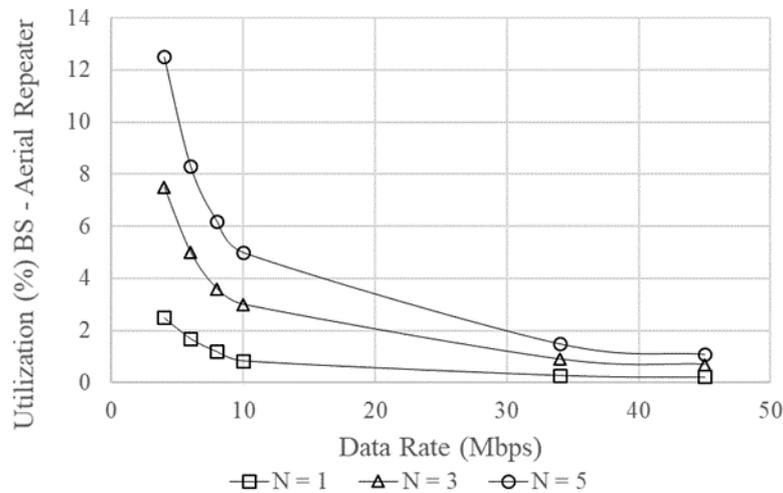

**Figure 5**: Dependences of "Ground Base Station - Aerial Repeater" channel *Utilization* on Transaction Size for different *Data Rates* (*TS* = 100 Kbits, *T* = 1 s)

### 3.4. Dependences of "Ground Base Station - Aerial Repeater" channel Utilization on BER

Let us consider the case when 10 Kbits packets are transmitted from a ground base station to users of a ground network at *Data Rate* of 45 Mbps. We will obtain graphs of channel *Utilization* on *BER* using the formula:

$$Utilization = ((TS \times N)/(Data\ Rate \times (1-BER) \times T))) \times 100\%. \qquad (3)$$

The code for Utilization calculation is shown in Appendix 4. The data in Figure 6 show how the channel load changes with increasing *BER*. As *BER* increases, the channel begins to operate less efficiently, since the percentage of erroneous bits increases, which leads to an increase in the load due to retransmissions.

The results allow us to estimate how stable the channel operates at different *BER* levels. High *BER* requires more resources for retransmission of data, which can lead to channel congestion, a decrease in speed and quality of communication.

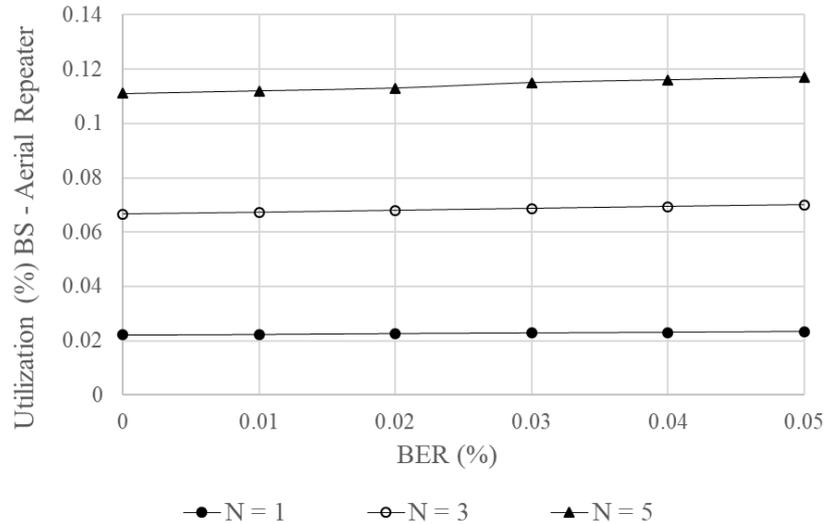

**Figure 6**: Dependences of "Ground Base Station - Aerial Repeater" channel *Utilization* on *BER*
(*Data Rate* = 45 Mbps, *TS* = 10 Kbits, *T* = 1 s)

### 4. AI-Assisted Adaptive Data Transfer

To implement adaptive control of transmission parameters, a lightweight supervised machine learning approach based on linear regression is employed. Linear regression is selected due to its low computational complexity, fast convergence, and suitability for real-time implementation on resource-constrained UAV platforms.

The objective of the AI model is to predict an appropriate transaction size based on observed latency and data rate, ensuring that latency remains below a predefined threshold. Training data are generated synthetically using simulation results obtained for different transaction sizes and data rates.

The linear regression model is trained to estimate the transaction size as a function of latency and data rate. Once trained, the model is used in an adaptive transmission loop, where the packet size is dynamically adjusted during data transfer. When the latency threshold is reached, the transaction size is reduced, and the adaptation process continues until stable operation is achieved.

Let us consider the following example of adaptation. The task is to create and train AI model using the dependencies between the delay $L$ and the packet size *TS* at different Data Rates. Based on this model, the adaptation of data transmission parameters will be performed, including changing the packet size and tracking the specified delay threshold.

In Figure 3 the X-axis is on a logarithmic scale for better visualization of the range of packet values. If we switch from a logarithmic to a regular scale, the dependencies for latency $L$ on packet size take the following form:

$$L_1 = 0.0005 \times TS \text{ for } Data\ Rate = 6 \text{ Mbps},$$
$$L_2 = 0.0003 \times TS \text{ for } Data\ Rate = 10 \text{ Mbps},$$
$$L_3 = 0.00007 \times TS \text{ for } Data\ Rate = 45 \text{ Mbps}.$$

To train the linear regression model, a data array is created using the $L_1, L_2, L_3$ dependencies with a *TS* step of 100 bits. The linear regression model is considered as a basic example of AI. Based on the created data array, a linear regression model is trained to predict the size of the transmitted *TS* packet for the specified *L* and *Data Rate*. The trained model is used to predict *TS* for *L* = 2 ms and *Data Rate* = 6 Mbps. The values of *TS, L* and *Data Rate* are printed.

In Python, the program code for adapting the transmission parameters is as follows:
- the predicted value of the size of the transmitted *TS* packet is rounded and then reduced by 1000 bits;
- data are transmitted, starting from the obtained *TS* value, increasing in 100-bit increments every minute of transmission;
- the transmission process is displayed on the *L(t)* dependence graph;
- as soon as the value of the y parameter reaches *L* = 2 ms, the *TS* value is printed, indicating that the adaptation threshold has been reached;
- after this, *TS* decreases by 400 bits and the transmission process continues, with a step of 100 bits every minute of transmission, displaying the *L(t)* dependence on the same graph;
- after reaching the value *L* = 2 ms three times, data transmission stops, and the channel parameter for adaptation procedure is displayed on the graphs.

The code for executing all steps of the task is shown in Appendix 5, and the calculated *L(t)* and *TS(t)* dependences are shown in Figures 7 and 8 respectively. These figures present the temporal evolution of latency and transaction size during the adaptive process. The results demonstrate that the proposed AI-assisted approach enables controlled adaptation of transmission parameters, preventing excessive latency growth under varying channel conditions.

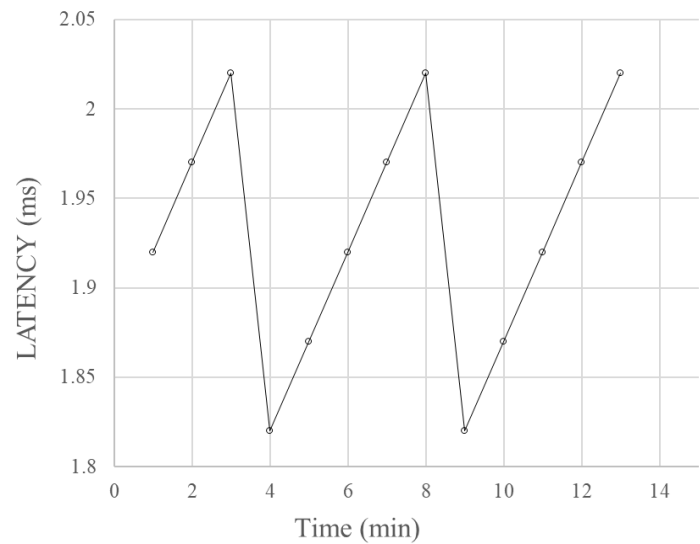

**Figure 7**: Graph of the dependence *L(t)*

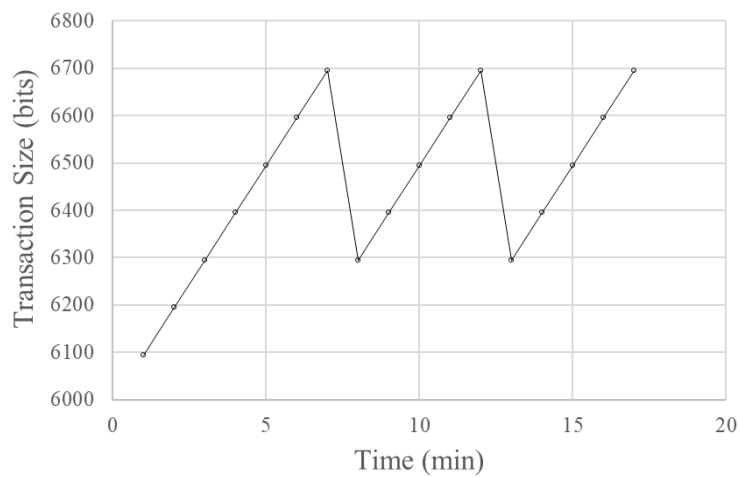

**Figure 8**: Graph of the dependence *TS(t)*

## 5. Conclusions

This paper presented a simulation-based study of AI-assisted communication channel adaptation in a UAV-enabled cellular network with clustered ground users. A lightweight supervised learning model based on linear regression was applied to adapt transaction size in response to changing channel conditions.

Simulation results demonstrated that adaptive control of transmission parameters improves latency stability and average channel utilization, particularly under high-interference conditions. The proposed approach does not rely on complex machine learning models and is therefore suitable for implementation on UAV platforms with limited computational resources.

The study confirms that even simple AI-assisted adaptation mechanisms can provide measurable performance benefits in UAV-assisted cellular communication systems. The presented model and results can serve as a basis for further investigations involving more advanced learning techniques, additional channel parameters, or dynamic user clustering strategies.

Future work will focus on extending the proposed framework to include more complex machine learning models, such as deep learning, and on investigating the impact of user mobility and dynamic cluster formation on adaptive channel performance.

## Appendix 1
### Code for *Latency* calculation

```
import matplotlib.pyplot as plt
import pandas as pd
# Parameters
packet_sizes = [10, 100, 1000, 10000, 100000] # Packet size in bits
data_rate = 1.54e6 # Data rate in bits/sec (1.54 Mbps)
num_links = 3 # Number of links in the channel (BS -> Relay -> UAV -> User)
# Calculate the delay for each packet size
delays = [(packet_size / data_rate) * num_links * 1000 for packet_size in packet_sizes] # Delay in ms
# Create data table
data = {
"Packet Size (bits)": packet_sizes,
"Total Delay (ms)": delays
}
df = pd.DataFrame(data)
print("Data table for plotting the graph:")
print(df)
# Plotting the graph
plt.figure(figsize=(10, 6))
plt.plot(packet_sizes, delays, marker='o', color='b', linestyle='-')
plt.xlabel("Packet Size (bits)")
plt.ylabel("Total Delay (ms)")
plt.xscale("log") # Logarithmic scale for packet size
plt.title("Packet transmission delay as a function of packet size")
plt.grid(True)
plt.show()
```

## Appendix 2
### Code for *Utilization* calculation

```
import matplotlib.pyplot as plt
import pandas as pd
# Given parameters
packet_sizes = [10, 100, 1000, 10000, 100000] # Packet sizes in bits
users = [1, 3, 5]
# Number of users
channel_bandwidth = 1.54e6
# Channel bandwidth in bits/s (1.54 Mbps)
# Table to store data
data = {'Packet Size (bits)': packet_sizes}
for user_count in users:
```

```python
load_percent = [(packet_size * user_count / channel_bandwidth) * 100 for packet_size in packet_sizes]
data[f'Load (%) for {user_count} user(s)'] = load_percent
# Create a DataFrame for the table
df = pd.DataFrame(data)
print("Data table for plotting graphs:")
print(df)
# Plotting graphs
plt.figure(figsize=(10, 6))
for user_count in users:
    load_percent = [(packet_size * user_count / channel_bandwidth) * 100 for packet_size in packet_sizes]
    plt.plot(packet_sizes, load_percent, label=f"{user_count} user(s)")
# Plot settings
plt.xlabel("Packet size (bits)")
plt.ylabel("Channel load (%)")
plt.title("Channel load dependence on packet size and number of users")
plt.legend()
plt.xscale('log') # Logarithmic scale for clarity
plt.grid(True)
plt.show()
```

## Appendix 3
**Code for calculation of Utilization on data transfer rate**

```python
import matplotlib.pyplot as plt
import pandas as pd
# Given parameters
packet_size = 100000 # Packet size in bits (100 Kbits)
num_users = 5 # Number of users
data_rates = [1.544e6, 2.048e6, 4e6, 6e6, 10e6, 34e6, 45e6] # Data rates in bits/sec
# Calculate channel load for each rate
load_percentages = [(packet_size * num_users / rate) * 100 for rate in data_rates]
# Create data frame
data = {
"Data Rate (Mbps)": [rate / 1e6 for rate in data_rates],
"Channel Load (%)": load_percentages
}
df = pd.DataFrame(data)
print("Data frame for plotting a graph:")
print(df)
# Plotting a graph
plt.figure(figsize=(10, 6))
plt.plot(data["Data Rate (Mbps)"], data["Channel Load (%)"], marker='o', color='b', linestyle='-')
plt.xlabel("Data Transfer Rate (Mbps)")
plt.ylabel("Channel Load (%)")
plt.title("Dependence of Channel Load on Data Transfer Rate")
plt.grid(True)
plt.show()
```

## Appendix 4
**Code for calculation of Utilization on BER**

```python
import matplotlib.pyplot as plt
import pandas as pd
# Parameters
packet_size = 10000 # Packet size in bits (10K)
num_users = 5 # Number of users
data_rate = 45e6 # Data rate in bits/sec (45M)
ber_values = [0, 0.01, 0.02, 0.03, 0.04, 0.05] # Bit error rates in %
# Calculate channel load for each BER value
channel_loads = [(packet_size * num_users / (data_rate * (1 - ber))) * 100 for ber in ber_values]
# Create data table
data = {
"BER (%)": ber_values,
```

```python
"Channel Load (%)": channel_loads
}
df = pd.DataFrame(data)
print("Data table for plotting the graph:")
print(df)
# Plotting the graph
plt.figure(figsize=(10, 6))
plt.plot(data["BER (%)"], data["Channel Load (%)"], marker='o', color='b', linestyle='-')
plt.xlabel("BER (%)")
plt.ylabel("Channel Load (%)")
plt.title("Dependence of channel load on the bit error rate (BER)")
plt.grid(True)
plt.show()
```

**Appendix 5**
**Code for adaptive data transfer modelling**

```python
import numpy as np
import pandas as pd
from sklearn.linear_model import LinearRegression
import matplotlib.pyplot as plt
import time
# Dependency parameters
data_rates = [6, 10, 45] # Data Rates in Mbps
slopes = [0.0005, 0.0003, 0.00007] # Slope coefficients for lag y vs batch size d
# Create training data
data = []
for rate, slope in zip(data_rates, slopes):
    for d in range(100, 10000, 100): # Batch size with 100-bit increments
        y = slope * d # Calculate lag
        data.append([y, rate, d])
# Convert data to DataFrame for model training
df = pd.DataFrame(data, columns=["y", "DataRate", "d"])
# Split into features and target variable
X = df[["y", "DataRate"]]
y = df["d"]
# Train linear regression model
model = LinearRegression()
model.fit(X, y)
# Predict packet size d for y = 2 ms and Data Rate = 6 Mbps
y_test = 2
data_rate_test = 6
predicted_d = model.predict([[y_test, data_rate_test]])
predicted_d = int(round(predicted_d[0])) # Rounding
# Adapt transmission parameters
d = predicted_d - 1000 # Decrease by 1000 bits
y_values = []
d_values = []
time_values = []
adaptation_count = 0
print(f"Initial packet size: {d} bits")
while adaptation_count < 3:
    # Calculate latency for current packet size
    y = 0.0005 * d if data_rate_test == 6 else (0.0003 * d if data_rate_test == 10 else 0.00007 * d)
    # Write data for graphs
    time_values.append(len(time_values))
    y_values.append(y)
    d_values.append(d)
    # Print information when adaptation threshold is reached
    if y >= 2:
        print(f"Adaptation threshold reached: latency {y:.2f} ms with packet size {d} bits")
        adaptation_count += 1
        d -= 400 # Reduce packet size after threshold is reached
```

```
else:
    d += 100 # Increasing the packet size
time.sleep(1) # Delay for data transfer simulation (1 minute for each step)
# Plotting y(t) and d(t) dependencies
fig, axs = plt.subplots(2, 1, figsize=(10, 8))
# Plotting y(t) dependency
axs[0].plot(time_values, y_values, marker="o", color="b")
axs[0].set_title("Dependence of y(t)")
axs[0].set_xlabel("Time (minutes)")
axs[0].set_ylabel("Delay y (ms)")
# Plotting d(t) dependency
axs[1].plot(time_values, d_values, marker="o", color="g")
axs[1].set_title("Dependence of packet size d(t)")
axs[1].set_xlabel("Time (minutes)")
axs[1].set_ylabel("Packet size d (bits)")
plt.tight_layout()
plt.show()
# Output data for building a table
table_data = pd.DataFrame({"Time (minutes)": time_values, "Delay y (ms)": y_values, "Packet   size d (bits)": d_values})
print("\nData table for graphs:")
print(table_data)
```

**Statements & Declarations**

**Funding**
The authors declare that no funds, grants, or other support were received during the preparation of this manuscript.
**Competing Interests**
The authors have no relevant financial or non-financial interests to disclose.
**Conflicts of Interest**
The authors declare no conflict of interest.
**Author Contributions**
Volodymyr Kharchenko – V.Kh., Andrii Grekhov – A.G., Vasyl Kondratiuk – V.K.
*Conceptualization*, A.G. and V.Kh.; *methodology*, A.G.; *validation*, A.G., V.Kh. and V.K.; *investigation*, A.G.; *resources*, V.Kh. and V.K.; *writing*—original draft preparation, A.G.; *writing*—review and editing,V.K.; *supervision*, V.Kh.; *project administration, V.K.*; All authors have read and agreed to the published version of the manuscript.
**Ethics approval**
Not applicable.
**Data Availability Statement**
All data generated and analyzed during this study are included in this article. The datasets generated during the current study are available from the corresponding author on request.